\documentclass[letterpaper, 11pt]{article}
\usepackage{amssymb,amsmath}
\usepackage{epsf}
\usepackage{times,bm,mathrsfs}  
\usepackage{amsmath}
\usepackage{amssymb}
\usepackage{amsfonts}
\usepackage{mathrsfs}
\usepackage{bbm}
\usepackage{color}
\usepackage{graphicx}
\usepackage{amscd}
\usepackage{epsfig}

\newcommand{\mysub}[1]{\raisebox{-0.5ex}{\scriptsize{#1}}}
\renewcommand{\thefootnote}{\fnsymbol{footnote}}

\author{
 Christel Kamp\footnote{Paul-Ehrlich-Institut, Federal Institute for Vaccines and Biomedicines, Paul-Ehrlich-Stra{\ss}e 51-59, 63225 Langen, Germany, kamch@pei.de}
}

\title{\vspace{0cm}
Untangling the interplay between epidemic spreading and transmission network dynamic}
\begin{document}
\date {}
\maketitle
\addtocounter{page}{-1}

\thispagestyle{empty}

\begin{abstract}

Epidemic spreading of infectious diseases is ubiquitous and has often considerable impact on public health and economic wealth. The large variability in spatio-temporal patterns of epidemics prohibits simple interventions and demands for a detailed analysis of each epidemic with respect to its infectious agent and the corresponding routes of transmission. 
To facilitate this analysis, a mathematical framework is introduced which links epidemic patterns to the topology and dynamics of the underlying transmission network. The evolution both in disease prevalence and transmission network topology are derived from a closed set of partial differential equations for infections without recovery which are in excellent agreement with complementarily conducted agent based simulations. 
The mutual influence between the epidemic process and its transmission network is shown by several case studies on HIV epidemics in synthetic populations. They reveal context dependent key processes which drive the epidemic and which in turn can be exploited for targeted intervention strategies.
The mathematical framework provides a capable toolbox to analyze epidemics from first principles. This allows for fast {\em in silico} modeling - and manipulation - of epidemics which is especially powerful if complemented with adequate empirical data for parametrization.

\vspace{1cm}

\end{abstract}

\clearpage

In spite of huge efforts to improve public health the spreading of infectious diseases is ubiquitous at the beginning of the 21st century with considerable variability in epidemic patterns among locations. Although the current influenza pandemic is a global challenge there are nonetheless differences in its timing in the northern and southern hemisphere due to seasonal effects \cite{ref167, ref150}. Another prominent example for epidemic variability is the prevalence of sexually transmitted diseases (STDs) and specifically HIV infections. While HIV is in many populations endemic at low levels or restricted to high risk groups it has become pandemic in other parts of the world \cite{ref168, ref191}. In consequence, the spreading of infectious diseases cannot be understood globally but is always the result of many local factors such as climatic and hygienic conditions, population density and structure, cultural habits and mobility - to name only a few  which in themselves are often interconnected and time dependent. 
Epidemic models aim to capture the mechanisms that link these factors to the epidemics that one eventually observes and to promote an understanding of the underlying dynamic processes as a prerequisite for intervention strategies \cite{ref43, ref44}. A useful abstraction in this context is to regard persons which may be infected as nodes of a network in which the links are the potentially infectious contacts among persons (or nodes in network notation). This simplification reveals one major cause for the large variability in epidemic processes: The structure and dynamics of this transmission network is highly context dependent: it depends on the routes of transmission and infectious profile of the pathogen, the behavioral patterns of its host as well as a number of other co-factors.

It is therefore important to develop models that can be flexibly adapted to specific epidemic situations. As our focus will be to study the interplay between transmission network topology and epidemics, we will restrict ourselves to diseases caused by agents which lead either to immunity or death in their host, i.e. in which infection can occur only once. These epidemics can be described by Susceptible-Infected-Recovered or SIR models \cite{ref43, ref44}. Whereas we refer to the mathematically closely related case in which an infection eventually leads to the death of the host as SID model (Susceptible-Infected-Death).
The classical SIR model describes epidemics in homogeneous and well mixed populations which is not generally a good approximation to the real world situation. Therefore, current epidemic models strive for a better consideration of the transmission network structure in order to make more realistic predictions \cite{ref52, ref117}. Compartmental models consider different contact patterns in sub-populations and link them via a contact matrix \cite{ref57}. Network approaches go further in directly considering the distribution in each person's number of infectious contacts $k$ (i.e. each node's degree $k$ in network notation) in the transmission network \cite{ref46, ref174,ref172}. These models allow to study transmission networks with strong heterogeneity in the number of contacts among persons which in some cases also consider correlations in the way contacts are made \cite{ref176,ref177} or clustering \cite{ref175}.  While these approaches focus on static networks, a recent approach considers networks with arbitrary degree distributions and transient contacts and allows to derive the number of susceptible and infected nodes from a closed set of equations  \cite{ref53,ref54}. Pair models are a very general approach \cite{ref178} to consider epidemics on heterogeneous networks which provide a lot of flexibility to consider correlations or clustering in the way contacts are made \cite{ref204,ref173} but as a trade-off they quickly become computationally very demanding \cite{ref169}.

Another assumption often implicitly made in epidemic models is that the epidemic sweeps through the population at time scales much shorter than the time scale of the background demographic process, i.e. natural birth and death processes are neglected. This can be considered a good approximation in cases such as the yearly influenza epidemics but it is hardly adequate for HIV epidemics which span decades. Taking motivation out of this shortcoming we further develop recent network epidemic models \cite{ref53,ref54} to consider demographic background processes. 
Having HIV as a case study in mind we focus on epidemics of diseases that lead to death after infection of susceptible persons, possibly after undergoing several stages of disease. Different from earlier work, our approach will also allow for an in detail study of the interplay between epidemic spreading and the structure and dynamics of the underlying transmission network.

\section{Introduction of the model}
We will first introduce the mathematical framework for a simple SID model with one infectious stage I before death D and study its basic properties. This will then be extended to a more general SI\protect\mysub{1}I\protect\mysub{2}D model with two stages of disease I\mysub{1} and I\mysub{2} before death.
\subsection{The SID model}
Following the ansatz of \cite{ref53,ref54}, heterogeneity in the number of infectious contacts of each person is introduced by splitting the population into subgroups according to their number of infectious contacts. 
For each of these subgroups evolution equations can be written down that take into account the effects induced both by the epidemic and the demographic background processes. 

With the notation of Tab. \ref{notationtab} the equations for the numbers of susceptible and infected persons with $k$ contacts $S_k$ and $I_k$ read:

\renewcommand{\thefootnote}{\fnsymbol{footnote}}
\begin{tabular}{clp{8cm}}
$\dot{S}_k=$ & $-rp_{SI}kS_k$& new infections\\
&$+\eta N\bar{p}_k-\eta S_k$ & natural birth and death\\
&$+\eta \bar{g}'(1,t)(S_{k-1}-S_k)$ & contacts made with new nodes\\
&$-\eta (kS_{k}-(k+1)S_{k+1})$ & contacts lost from dying nodes\\
&$-\mu p_{SI} (kS_{k}-(k+1)S_{k+1})$ & contacts lost from nodes dying from infection\\  
$\dot{I}_k=$ & +$rp_{SI}kS_k$& new infections\\
& $-(\eta+\mu) I_k$ & death\\
&$+\eta \bar{g}'(1,t)(I_{k-1}-I_k)$ & contacts made with new nodes\\
&$-\eta (kI_{k}-(k+1)I_{k+1})$ & contacts lost from dying nodes\\
&$-\mu p_{II} (kI_{k}-(k+1)I_{k+1})$ & contacts lost from nodes dying from infection\\  
\end{tabular}

\begin{table}[h!]
\begin{center}
\caption{Notation of the model*
}\label{notationtab}
\begin{tabular}{@{\vrule height 10.5pt depth4pt  width0pt}lp{8cm}}
\hline
$A_k$ ( $A=\sum_k A_k$ )& number of persons in group $A$ with $k$ contacts (and total no. in $A$)\\
$N_k=\sum_A A_k$ ($N=\sum_k N_k$ )& number of persons with $k$ contacts (and total no. of persons) \\
$p_{Ak}=\frac{A_k}{A}=\frac{g^{(k)}_A(0,t)}{k!}$& probability for a person in group $A$ to have $k$ contacts\\
$g_A(x,t) = \sum_k p_{Ak}(t)x^k$ & probability generating function (PGF) of $p_{Ak}(t)$ \\
$\langle k\rangle_A = g'_A(1,t)$ & average number of contacts of $A$ persons \\
$p_{k}=\frac{N_k}{N}$& probability for a person to have $k$ contacts\\
$g(x,t) = \sum_k p_{k}(t)x^k=\sum_A \frac{A}{N}g_A(x,t)$ & probability generating function (PGF) of $p_{k}(t)$ \\
$\langle k\rangle= g'(1,t)$ & average number of contacts \\
$\bar{p}_{k}$& probability for a person entering the population to have $k$ contacts\\
$\bar{g}(x,t) = \sum_k \bar{p}_{k}(t)x^k$ & probability generating function (PGF) of $\bar{p}_{k}(t)$ \\
$M_A=\sum_k k A_k=Ag'_A(1,t)$  ( $M =\sum_A M_A$)& number of links coming from $A$ persons (and total no. of links)\\
$M_{AB}$ & number of links coming from $A$ persons and pointing to $B$ persons\\
$p_{AB}=\frac{M_{AB}}{M_A}$ & probability for a link starting form an $A$ person to point to a $B$ person\\ 
\hline
\multicolumn{2}{p{15.5cm}}{*Note that $A$, $B$ correspond to the stages that are passed during an infection, e.g. $S$ for susceptible, $I$ for infected etc. Throughout the manuscript derivatives with respect to time/spatial variables are denoted by a dot/prime.}\\
\hline
\end{tabular}
\end{center}
\end{table}

Note that implicitly is assumed that individuals who newly enter the population have $k$ contacts with probability $\bar{p}_k$ and link randomly to those persons already present. People dying from natural causes are assumed to have the same average number of contacts as found in the whole population without preferences for susceptible or infected persons. 
In terms of the total number of susceptible $S$ and infected persons $I$ the equations read
\begin{eqnarray}
\dot{S} &=& \eta N -rp_{SI} M_S - \eta S \label{si1r}\\
\dot{I} &=& rp_{SI} M_S - (\eta+\mu) I.
\end{eqnarray}
To close this set of equations additional equations have to be derived for $p_{SI}$ and $p_{II}$, the probabilities that a contact made by a susceptible or infected person points to an infected person. To capture the evolution of the network topological features the time evolution of the probability generating functions (PGF) $g_S(x,t)$ and $g_I(x,t)$ have to be calculated which represent the degree distributions of susceptible and infected persons. That means we have to follow the way susceptible and infected persons are linked and how the topological features of the transmission network change within these subgroups in response to the epidemic.
To derive $p_{SI}$ and $p_{II}$ we follow the techniques developed in \cite{ref53} (cf. supporting information for detailed calculations) and conclude
\begin{eqnarray}
\dot{p}_{SI} &=& r\frac{g''_S(1,t)}{g'_S(1,t)}p_{SI}(1-p_{SI})-(r+\mu)p_{SI}(1-p_{SI})\nonumber\\
&&+\eta\frac{\bar{g}'(1,t)}{M_S}(I-(N+S)p_{SI})\label{psi}\\
\dot{p}_{II} &=& r\frac{M_S}{M_I}\frac{g''_S(1,t)}{g'_S(1,t)}p_{SI}(2p_{SI}-p_{II}) -r\frac{M_S}{M_I}p_{SI}p_{II}\nonumber\\
&&+2r\frac{M_S}{M_I}p_{SI}- \mu (1-p_{II})p_{II} +\eta\frac{\bar{g}'(1,t)} {M_I}Ip_{II}.\label{pii}      
\end{eqnarray}
The occurrence of the probability generating functions in equations (\ref{psi}) and (\ref{pii}) makes obvious that the way links are maintained between susceptible and infected nodes depends on the contact behavior within these subgroups as well as the changes therein, or in other words, the time evolution of the PGFs $g_S(x,t)$ and $g_I(x,t)$ (via $g'_S(1,t)$, $g''_S(1,t)$ and $M_A=Ag'_A(1,t)$, $A\in\{S,I\}$). Considering $\dot{g}_A(x,t)=\sum_k\left(\frac{\dot{A}_k}{A}-\frac{\dot{A}}{A}p_{Ak}\right)x^k$
the set of equations can be closed with
\begin{eqnarray}
\dot{g}_S(x,t)&=&-rp_{SI}(xg'_S(x,t)-g'_S(1,t)g_S(x,t))\nonumber\\
&&+\eta\frac{N}{S}(\bar{g}(x,t)-g_S(x,t))\nonumber\\
&&-\eta(1-x)\bar{g}'(1,t)g_S(x,t)\nonumber\\
&&+(\eta+\mu p_{SI})(1-x)g'_S(x,t)\\
\dot{g}_I(x,t)&=& r p_{SI}\frac{S}{I}(xg'_S(x,t)-g'_S(1,t)g_I(x,t))\nonumber\\
&&-\eta(1-x)\bar{g}'(1,t)g_I(x,t)\nonumber\\
&&+(\eta+\mu p_{II})(1-x)g'_I(x,t).\label{sig}
\end{eqnarray}
With the equations given in Tab. \ref{notationtab} we are able to investigate the interplay between epidemic processes and  transmission network topology. In particular, it is possible to follow the full degree distribution within the subgroups of susceptible and infected persons in terms of the probability generating functions $g_S(x,t)$ and $g_I(x,t)$. 
However, transmission networks do not only change their structure as a result of birth and death processes but also due to changes in contact partners. Analogous to \cite{ref54} we consider swapping of contact partners at a rate $\rho$ which affects the quantities $p_{SI}$ and $p_{II}$ via an additional term $\rho\left(\frac{M_I}{M}-p_{SI}\right)$ and $\rho\left(\frac{M_I}{M}-p_{II}\right)$ in equations (\ref{psi}) and (\ref{pii}), respectively.

\begin{figure*}
\begin{center}
\includegraphics[angle=0, height = 7.8cm]{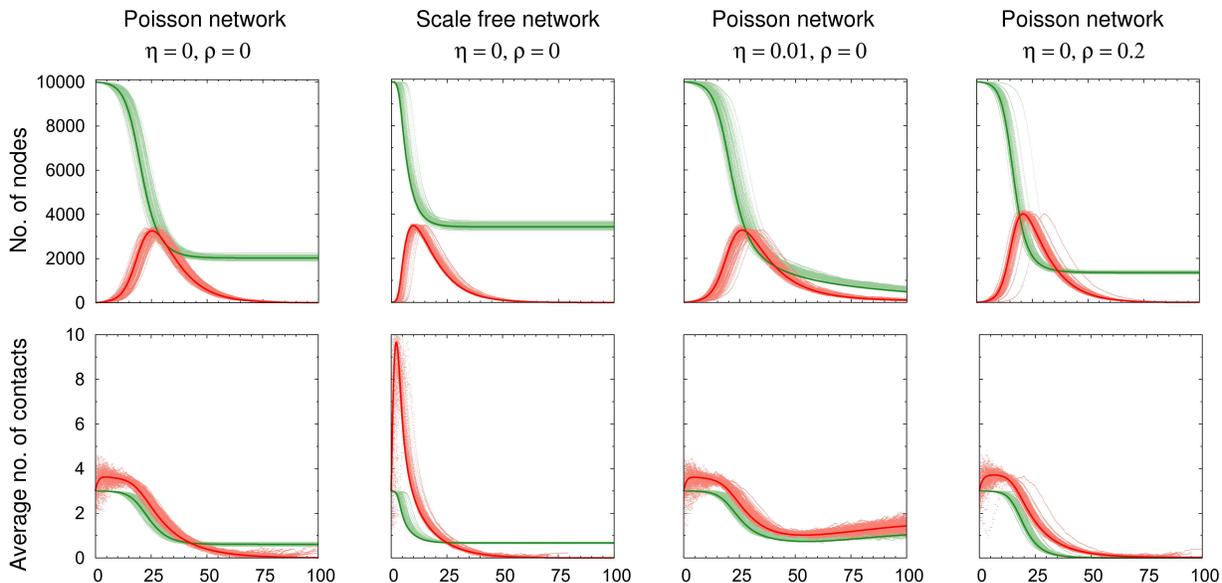}
\caption{Evolution in the numbers of susceptible (green) and infected (red) persons (top panel) as well as their average number of contacts per person (bottom panel). The doted, light colored curves correspond to the result of 100 agent based simulations, the solid lines are the results of the numerical solution of the set of partial differential equations (\ref{si1r}-\ref{sig}), parameters are chosen analogous to \protect{\cite{ref53}}, average number of contacts $\langle k\rangle$ = 3, transmission rate $r$ = 0.2, recovery rate $\mu$ = 0.1. Epidemics in networks with differences in heterogeneity and transience in contacts are shown: The first column shows a static Poisson network with degree distribution $p_k=\frac{\langle k\rangle^ke^{-{\langle k \rangle}}}{k!}$ as opposed to a network with a scale free degree distribution $p_k=\frac{k^{-\gamma}e^{-\frac{k}{\kappa}}}{Li_{\gamma}\left(e^{-\frac{1}{\kappa}}\right)}$ in column 2 (same average degree), column 3 corresponds to the network in column 1 with an additional demographic process (birth and death at a rate $\eta=0.01$), column 4 corresponds to column 1 with the additional feature that contact partners change at a rate $\rho=0.2$. }\label{sid}
\end{center}
\end{figure*}

This approach allows to study and analyze epidemics on transmission networks with a broad range of topological features. In particular, any degree distribution  which represents the contact behavior within a population can be implemented within the formalism by providing the corresponding probability generating function as an input parameter.
Figure \ref{sid} shows the excellent agreement between the predictions derived from the set of partial differential equations (\ref{si1r}-\ref{sig}) and the observations from agent based simulations for several exemplary topologies. The approach reproduces in particular well known behavior: Epidemics in heterogeneous networks spread faster but are more restricted (Fig. \ref{sid}, column 1 vs. 2) \cite{ref197}. Contrarily, transient contacts lead to a wider spread of disease as long as contacts are maintained sufficiently long to ensure transmission (Fig. \ref{sid}, column 1 vs. 4). The re-growth in the average number of contacts per person (node) in the scenario with birth and death processes (Fig. \ref{sid}, column 3) is a reminiscent of re-emergent or persistent infections that can be observed in this context. 

\subsection{The SI\protect\mysub{1}I\protect\mysub{2}D model}
Following the intention to model HIV epidemics we have to extend the model to take the heterogeneous infectious profile of an HIV infection into account. The course of disease is characterized by a short but highly infectious period of primary infection. This is followed by a prolonged period of latent infection with a much lower infectivity (sometimes also referred to as asymptomatic or chronic phase) \cite{ref76} before the onset of AIDS as a final stage of disease.
In order to understand HIV epidemiology it is therefore important to consider the different stages of disease. Our focus will be on the primary infection $I_1$ which ceases at a rate $\mu_1$ and is associated with a transmission rate $r_1$ as opposed to the latent infection $I_2$ with parameters $\mu_2$ and $r_2$. We neglect the role of the final stage of disease in transmission assuming that health conditions will be in conflict with a further transmission of HIV. 
The evolution equations for $S_k$, $I_{1k}$ and $I_{2k}$ are determined analogously to the SID case to describe the numbers of person with $k$ contacts being susceptible, primarily infected and latently infected. In addition, equations describing the contacts among persons of different epidemic groups are derived (i.e. $p_{SI_1}$, $p_{SI_2}$, $p_{I_1I_1}$, $p_{I_1I_2}$ and $p_{I_2 I_2}$). Finally the set of equations is closed by a derivation of the probability generating functions $g_S(x,t)$, $g_{I_1}(x,t)$ and $g_{I_2}(x,t)$ which describe the contact patterns and their temporal changes within each group (for details cf. supporting information).
The resulting equations may look cumbersome but allow for a much faster (and flexible) assessment of epidemic scenarios than agent based simulations. Again it becomes clear in the mathematical structure of the equations that the epidemic process cannot be decoupled from the underlying transmission network and that there is a mutual influence between both. 

\begin{figure*}
\begin{center}
\includegraphics[angle=0, height=4cm]{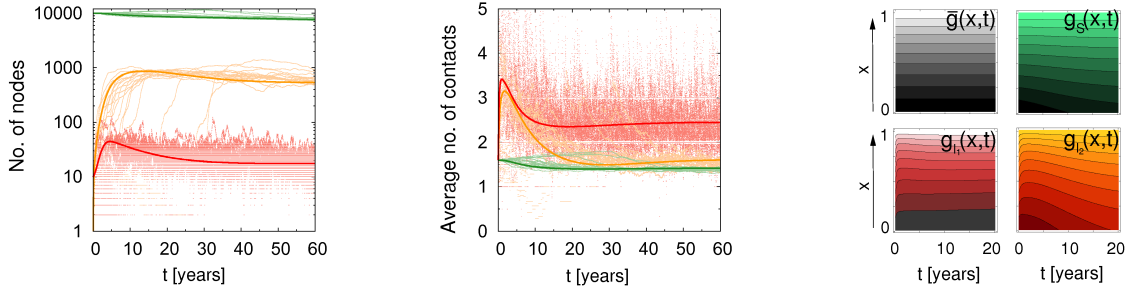}
\caption{Evolution in the numbers of susceptible (green), primarily (red) and latently (orange) infected persons (left) as well as their average number of contacts per person during the epidemic (middle). The doted, light colored curves correspond to the result of 20 agent based simulations, the solid lines are the results of the numerical solution of the set of partial differential equations. Note that logarithmic scale was chosen to present the different orders of magnitude in the size of the epidemic subgroups. The epidemics takes place on a scale free network with average degree $\langle k \rangle=1.6$ into which persons are born at a rate $\eta_1=0.02a^{-1}$ and die at a rate $\eta_2=0.015a^{-1}$.
The epidemic parameters are chosen in accordance with the infectious profile of HIV \protect\cite{ref76}, i.e. transmission rates in the stages of primary and latent infection are $r_1=2.76a^{-1}$ and $r_2=0.1a^{-1}$ with rates of progression $\mu_1=4.1a^{-1}$ and $\mu_2=0.12a^{-1}$.
The right panel shows the evolution of the probability generating functions over 20 years of the epidemic representing the distribution in the number of contacts in persons newly entering the population ($\bar{g}(x,t)=const.$) as well as in susceptible, primarily and latently infected persons. The contour plots interpolate from 0 (dark colors) to 1 (light colors) in steps of 0.1.}\label{si12d}
\end{center}
\end{figure*}

This is exemplarily illustrated in Fig. \ref{si12d} which shows the spreading of HIV in a ``synthetic" population with a scale free distribution in the number of potentially infectious contacts. ($p_k=\frac{k^{-\gamma}e^{-\frac{k}{\kappa}}}{Li_{\gamma}\left(e^{-\frac{1}{\kappa}}\right)}$, $\gamma=2.4$, $\kappa=20$, average number of contacts $\langle k\rangle =1.6$) to which persons are born at a rate of $\eta_1=0.02$ per year and die at a rate  $\eta_2=0.015$ per year. Transmission and progression rates of HIV in the primary and latent stage of disease $r_1$, $r_2$, $\mu_1$ and $\mu_2$ are estimated from \cite{ref76}.
Agent based simulations and the numerical solution of the set of partial differential equations agree well and show how the epidemic saturates within a few decades with a few percent of latently infected individuals (as opposed to a few per mill of primarily infected individuals). During the epidemic expansion phase the average number of contacts among infected persons grows sharply while the average number of contacts among susceptible persons decreases slightly. Persons with more potentially infectious contacts are at a higher risk of infection and accumulate within the population of infected persons. This hierarchy in the average number of contacts from primarily infected over latently infected to susceptible individuals can still be observed during the saturation phase of the epidemic.
The temporal evolution in the network topology can be studied in more detail from the probability generating functions $g_S(x,t)$, $g_{I_1}(x,t)$ and $g_{I_2}(x,t)$ of the degree distributions describing the contact behavior within the epidemic subgroups in Fig. \ref{si12d} (right panel).
The contact behavior of individuals newly entering the population is not time dependent and corresponds to original distribution found before the onset of the epidemic ($\bar{g}(x,t)=g(x,0)=const$). Nonetheless, the contact patterns in the population change specifically in the epidemic subgroups. Considering that $g_A(0,t)=p_{A0}$ is the probability to find a person without infectious contacts in each group one can see that the fraction of single persons grows among susceptible persons as it does among latently infected persons in the saturation phase of disease due to the increased death rate among their infected contacts. This loss in contacts is not observed in the short period of primary infection. It corresponds to the maintenance of the high average number of contacts in this subgroup (originally acquired due to the hazard of infection growing with the number of contacts). The full probability distribution for the observed numbers of contacts per person in each group can be derived from the shown probability generating functions $g_A(x,t)$ via $p_{Ak}(t)=\frac{g^{(k)}_A(0,t)}{k!}$. The changes in structure of the PGFs among the subgroups become even more obvious if we start with a more homogeneous topology as for example a network with a Poissonian degree distribution.

Note that a persistent epidemic would not be equally observed if new persons enter the population with the currently found contact behavior instead of the initially found contact behavior, i.e. with $\bar{g}(x,t)=g(x,t)$ instead of $\bar{g}(x,t)=g(x,0)=const$. As high risk individuals have a higher risk of death due to the epidemic this will lead successively to an introduction of individuals with lower risk behavior (lower average number of contacts) until eventually the network becomes sub-critical and the epidemic ceases.

\section{Transmission by stage of disease}
The analysis has already taken into account that the course of an HIV infection is intimately related to the dynamics of its epidemic spreading: While the brief period of primary infection is associated with a largely increased infectivity one observes a much lower infectivity during the prolonged period of latent infection (sometimes also referred to as asymptomatic or chronic infection) which grows again in the late stages of disease \cite{ref76,ref179, ref183, ref186}. Therefore it cannot trivially be decided from which phase of disease most new infections emanate making this a topic of ongoing debate \cite{ref189,ref188}. 
There are various studies on the contribution of the initial and latent stage to HIV incidence in different populations (i.e. in largely heterosexual populations \cite{ref180} and homosexual populations \cite{ref181}). They reveal that the contribution of either stage of disease to HIV incidence is very context dependent. There is however agreement that primary infection becomes a more important driver of the epidemic when risk behavior increases (number of casual/concurrent contacts or partners) whereas incidence from latent infections becomes more important in the saturation phase of an epidemic (as opposed to its expansion phase) which is supported by modeling approaches \cite{ref182,ref184}. 

\begin{figure*}
\begin{center}
\includegraphics[angle=0,height=5cm]{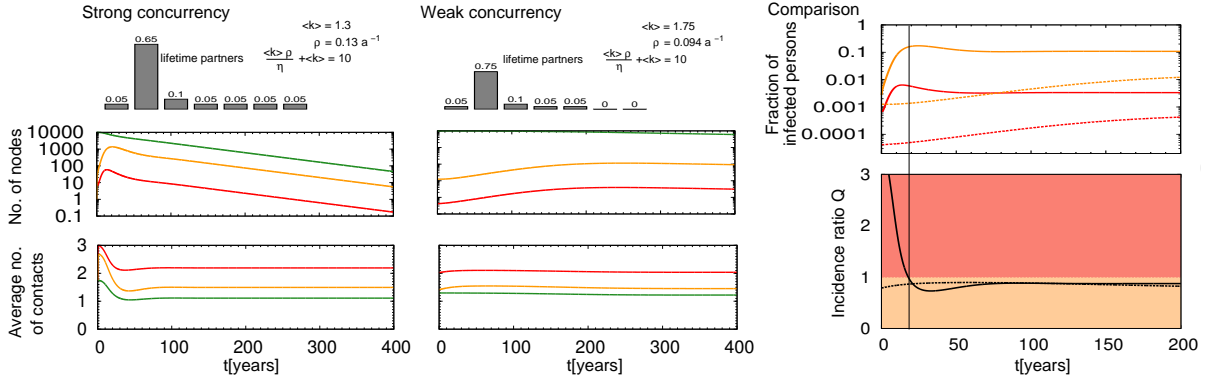}
\caption{Evolution in the numbers of susceptible (green), primarily (red) and latently (orange) infected persons as well as their average number of contacts per person in the scenario of strong and weak concurrency (left and middle column). Note that logarithmic scale was chosen to present the different orders of magnitude in the size of the epidemic subgroups. The epidemics take place on random networks with the sketched distributions in the number of contacts ($p_0$ to $p_6$ are shown, $p_k=0$ for $k>6$) into which persons are born and die at a rate $\eta_1=\eta_2=0.02a^{-1}$.
The epidemic parameters are chosen in accordance with the infectious profile of HIV \protect\cite{ref76}, i.e. transmission rates in the stages of primary and latent infection are $r_1=2.76a^{-1}$ and $r_2=0.1a^{-1}$ with rates of progression $\mu_1=4.1a^{-1}$ and $\mu_2=0.12a^{-1}$.
The right column (comparison, top) shows the fraction of persons in the primary (red) and latent (orange) stage of disease in the scenarios of strong concurrency (solid lines) and weak concurrency (dashed lines) for comparison. The final diagram shows the incidence ratio for both scenarios, i.e. the fraction of the number new infections from the primary over those from the latent stage of disease (strong concurrency - solid line, weak concurrency - dashed line).}\label{hclc}
\end{center}
\end{figure*}

This question can certainly not be answered conclusively by our study due to lack of adequate empirical data but we can provide a tool which helps to better understand under which circumstances infections dominate either from primarily infected or latently infected persons. Therefore we do not aim to parametrize the model in the most realistic way but investigate two possible scenarios which consider some general features found in HIV epidemic networks via sexual transmission. There is generally large heterogeneity in the numbers of sex partners \cite{ref49,ref47,ref77} as well as in the frequency of partner change and the level of concurrency in partnerships \cite{ref190}. Acknowledging that not only the number of infectious contacts but also their timing \cite{ref78,ref123,ref124} is important for epidemic spreading we investigate two scenarios that we depict as weak and strong concurrency with parameters as given in Fig. \ref{hclc}. Both scenarios assume an average number of 10 lifetime partners during a lifetime of 50 years ($\eta=0.02 a^{-1}$, neglecting delays due to childhood/adolescence before sexual debut). However, in the scenario of weak concurrency a lower average number of concurrent partners $\langle k \rangle$ is traded in for a higher partner change rate $\rho$ in comparison to the scenario of strong concurrency.

The results from the numerical solution of the set of PDEs is shown in Fig. \ref{hclc}. It gets immediately apparent that not only the number of partners but also the level of their concurrency has a profound impact on both the time scale and width of epidemic expansion. One observes a much more severe epidemic in the case of strong concurrency. A common feature of both epidemics is the hierarchy in the average number of contacts for primarily, latently infected and susceptible persons giving directions for targeted intervention strategies.
To understand which stage of disease drives the epidemics it is specifically of interest to study the ratio in the number of infections that are derived from individuals in their primary and latent stages of disease, i.e. the incidence ratio 
\begin{equation}
 Q=\frac{r_1p_{SI_1}}{r_2p_{SI_2}}
\end{equation}
which is shown in the bottom right panel of Fig. \ref{hclc}. While infections from the primary stage of disease dominate in the expansion phase of the epidemic in a scenario with strong concurrency in contacts, infections from the latent stage of disease dominate as soon as the epidemic matures. In the case of weak concurrency (being a closer approximation to serial monogamy) infections from the latent stage of disease dominate throughout the whole epidemic. This case study confirms that the question which stage of disease drives HIV epidemics cannot be answered without profound knowledge of the topology and dynamics of the underlying transmission network as well as knowledge about the saturation stage of the epidemic.

\section{Discussion}
A highly flexible mathematical framework has been introduced which allows to investigate the interplay between the topology and dynamics of the transmission network and the epidemics of a corresponding infectious agent within a closed set of equations. The approach has been focused on pathogens which lead to death of their hosts after some time of infection (possibly in several stages). HIV epidemics have been considered as an area of application in which the newly developed method helps to understand the complex interdependencies between the HIV epidemic profile, its transmission network, and epidemic process. Several scenarios in synthetic populations have been investigated showing that the transmission network is not a static support that shapes the epidemic but to the contrary is itself shaped by the epidemic. This becomes manifest in the changing contact behavior of infected and susceptible persons quantified in terms of the probability generating functions of their degree distributions. This interplay can become even more intricate in the presence of a multi-stage infectious profile as observed for HIV. Still, the mathematical framework provides a capable tool to derive context dependent answers to the question which stage of disease is the major driver of the epidemic.
Together this allows for a context sensitive identification of key epidemic processes and (sub-)groups that have to be targeted for maximum impact. The HIV case studies emphasize that it depends both on the maturation stage of the epidemics and the structure of the relevant transmission network on which groups intervention strategies (i.e. for example promotion of behavioral changes or vaccination programs if available) should focus.

The current presentation gives only one example for the application of the developed method which can be extended to other settings with a more complex infectious profile or to epidemics with a classical SIR dynamic. This makes the approach useful to a very broad spectrum of epidemic scenarios which may include improved modeling of SIR-like infections such as measles, rubella, pertussis or influenza.
The broad applicability of the approach makes it worthwhile to consider further improvements which stretch the limits of the model towards an increasingly realistic description of the epidemics we face day-to-day.
Fig. \ref{sid} and Fig. \ref{si12d} already indicate that the current approach is designed for the limit of infinite population sizes and does not account for the obvious fluctuations seen in the finite size agent based simulations. However, recent research has shown that stochastic fluctuations may have a strong influence on real world epidemic phenomena such as re-emergent epidemics \cite{ref199,ref200} making this an exciting direction of future research within the given mathematical framework.
It is further assumed that contacts are made randomly \cite{ref172} without taking into account any preferences or correlations influencing their establishment. Social networks usually show clustered communities \cite{ref202} and some degree of assortativity, i.e. persons tend to mix with their likes. \cite{ref203,ref176,ref57}. This often results in the generation of core groups which sustain and drive an epidemic and which will have to be considered in an extended model.

In conclusion, we have presented a mathematical framework that allows to closely monitor both the epidemic process and its transmission network for general SIR-like infections in an computationally efficient way. The current method allows for  great flexibility to account for variability in transmission network topology and dynamics as well as pathogen specific features. Nonetheless, it will be important to assess the method's limitations in the field after parametrization with appropriate empirical data - and an exciting challenge for future research to expand its limits further. 
\section{Materials and Methods}
Agent based simulations were done with NetLogo V4.0.4 \cite{ref192}, in which some code fragments were used from the model ``Virus on Network'' which is included in the software's model library \cite{ref194}. Poisson networks were generated by assigning $\langle k \rangle N/2$ links between randomly picked nodes, other random networks were generated on the basis of their degree sequence \cite{ref201,ref172}.
The numerical solution of the partial differential equations was done with Mathematica V6.0.2 \cite{ref195} using the function NDSolve and the numerical method of lines \cite{ref196}. 

\section*{Acknowledgments}
I would like to thank D. Hollingsworth for the hospitality and inspiring discussions during my visit in the Department of Infectious Disease Epidemiology at Imperial College, London, J. L\"ower for the overall support of the project at the Paul-Ehrlich-Institut, and A. Bunten for his feedback on the manuscript.

\newpage
\section*{Supporting Information}

The following paragraphs provide a detailed derivation of the equations for the SID and SI\protect\mysub{1}I\protect\mysub{2}D models in which the notation introduced in Tab. \ref{notationtab} is used.

\subsection*{Equations for the SID model}

We consider an epidemic of a disease that is transmitted at a rate $r$ and from which infected persons die at a rate $\mu$. The demographics of the background population is determined by a birth rate $\eta_1$ and a death rate $\eta_2$. This leads to the following equation for the number of susceptible and infected persons with $k$ (infectious) contacts
\begin{eqnarray}
\dot{S}_k&=& -rp_{SI}kS_k+\eta_1 N\bar{p}_k-\eta_2 S_k+\eta_1 \bar{g}'(1,t)(S_{k-1}-S_k)-(\eta_2+\mu p_{SI}) (kS_{k}-(k+1)S_{k+1}) \\
\dot{I}_k&=&+rp_{SI}kS_k-(\eta_2+\mu) I_k+\eta_1 \bar{g}'(1,t)(I_{k-1}-I_k)-(\eta_2+\mu p_{II}) (kI_{k}-(k+1)I_{k+1}).
\end{eqnarray}

In terms of the total number of susceptible and infected persons the equations read
\begin{eqnarray}
\dot{S} &=& \eta_1 N -rp_{SI} M_S - \eta_2 S\\
\dot{I} &=& rp_{SI} M_S - (\eta_2+\mu) I.
\end{eqnarray}

To close this set of equations we also have to derive equations for $p_{SI}$ and $p_{II}$ as well as for the probability generating functions (PGF) $g_S(x,t)$ and $g_I(x,t)$. Following the argument in \cite{ref53} we write 
\begin{eqnarray}
\dot{p}_{SI} &=&\frac{\dot{M}_{SI}}{M_S}-\frac{\dot{M}_{S}}{M_S}p_{SI}\\
\dot{p}_{II} &=& \frac{\dot{M}_{II}}{M_I}-\frac{\dot{M}_{I}}{M_I}p_{II}
\end{eqnarray}
to derive equations for the number of links among susceptible and infected persons.
Noting that the average excess degree of a susceptible node that was reached from an infected node to susceptible or infected nodes are $\delta_{SI}(S)=(1-p_{SI})\frac{g''_S(1,t)}{g'_S(1,t)}$ and $\delta_{SI}(I)=p_{SI}\frac{g''_S(1,t)}{g'_S(1,t)}$, respectively, one can write
\begin{eqnarray}
\dot{M}_S&=& -rp_{SI}S(g''_S(1,t)+g'_S(1,t))+\eta_1\bar{g}'(1,t)(N+S)-(2\eta_2+\mu p_{SI})M_S\\
\dot{M}_I&=&=rp_{SI}S(g''_S(1,t)+g'_S(1,t))+\eta_1\bar{g}'(1,t)I-(2\eta_2+\mu+\mu p_{II})M_I\\
&\nonumber\\
\dot{M}_{SI}&=& -rp_{SI}M_S (\delta_{SI}(I)-\delta_{SI}(S))-(r+\mu)M_{SI}-2\eta_2 M_{SI}+\eta_1\bar{g}'(1,t)I\\
&=& -r(2p_{SI}-1)p_{SI}M_S \frac{g''_S(1,t)}{g'_S(1,t)}-(r+\mu+2\eta_2)M_{SI}+\eta_1\bar{g}'(1,t)I\\
\dot{M}_{II}&=& 2rp_{SI}M_S \delta_{SI}(I)+2rM_{SI}-2\mu M_{II}-2\eta_2 M_{II}\\
&=&2rp_{SI}^2M_S \frac{g''_S(1,t)}{g'_S(1,t)}+2rM_{SI}-2(\mu+\eta_2)M_{II}
\end{eqnarray}
and finally 
\begin{eqnarray}
\!\!\!\!\!\!\dot{p}_{SI} \!&=& \!rp_{SI}(1-p_{SI})\frac{g''_S(1,t)}{g'_S(1,t)}-(r+\mu)p_{SI}(1-p_{SI})+\eta_1\frac{\bar{g}'(1,t)}{M_S}(I-(N+S)p_{SI})\\
\!\!\!\!\!\!\dot{p}_{II} \!&=& \!r\frac{M_S}{M_I}p_{SI}(2p_{SI}-p_{II})\frac{g''_S(1,t)}{g'_S(1,t)} -r\frac{M_S}{M_I}p_{SI}p_{II}+2r\frac{M_S}{M_I}p_{SI}- \mu (1-p_{II})p_{II} +\eta_1\frac{\bar{g}'(1,t)} {M_I}Ip_{II}.    
\end{eqnarray}
To close the set of equations we have also to derive the probability generating functions (PGF) according to 
\begin{eqnarray}
\dot{g}_S(x,t)&=&\sum_k\left(\frac{\dot{S}_k}{S}-\frac{\dot{S}}{S}p_{Sk}\right)x^k\\
\dot{g}_I(x,t)&=&\sum_k\left(\frac{\dot{I}_k}{I}-\frac{\dot{I}}{I}p_{Ik}\right)x^k
\end{eqnarray}
which result in 
\begin{eqnarray}
\dot{g}_S(x,t)&=&-rp_{SI}(xg'_S(x,t)-g'_S(1,t)g_S(x,t))+\eta_1\frac{N}{S}(\bar{g}(x,t)-g_S(x,t))\nonumber\\
&&-\eta_1(1-x)\bar{g}'(1,t)g_S(x,t)+(\eta_2+\mu p_{SI})(1-x)g'_S(x,t)\\
\dot{g}_I(x,t)&=& r p_{SI}\frac{S}{I}(xg'_S(x,t)-g'_S(1,t)g_I(x,t))\nonumber\\
&&-\eta_1(1-x)\bar{g}'(1,t)g_I(x,t)+(\eta_2+\mu p_{II})(1-x)g'_I(x,t).
\end{eqnarray}

\subsection*{Equations for the SI\protect\mysub{1}I\protect\mysub{2}D model}
We extend the equations of the SID model to accommodate two infected stages $I_1$ and $I_2$ with transmission rates $r_1$ and $r_2$ which cease at rates $\mu_1$ and $\mu_2$, respectively.
The initial equations for the number of persons with $k$ contacts who are in the classes $S$, $I_1$ or $I_2$ read accordingly
\begin{eqnarray}
\dot{S}_k&=& -(r_1p_{SI_1}+r_2p_{SI_2})kS_k+\eta_1 N\bar{p}_k-\eta_2 S_k+\eta_1 \bar{g}'(1,t)(S_{k-1}-S_k)\nonumber\\
&&-(\eta_2+\mu_2 p_{SI_2}) (kS_{k}-(k+1)S_{k+1}) \\
\dot{I}_{1k}&=&+(r_1p_{SI_1}+r_2p_{SI_2})kS_k-(\eta_2+\mu_1) I_{1k}+\eta_1 \bar{g}'(1,t)(I_{1 k-1}-I_{1k})\nonumber\\
&&-(\eta_2+\mu_2 p_{I_1I_2}) (kI_{1k}-(k+1)I_{1 k+1})\\
\dot{I}_{2k}&=&\mu_1 I_{1k}- (\eta_2+\mu_2) I_{2k}+\eta_1 \bar{g}'(1,t)(I_{2 k-1}-I_{2k})\nonumber\\
&&-(\eta_2+\mu_2 p_{I_2I_2}) (kI_{2k}-(k+1)I_{2 k+1}).
\end{eqnarray}
In terms of the total number of susceptible and infected persons the equations read
\begin{eqnarray}
\dot{S} &=& \eta_1 N -(r_1p_{SI_1}+r_2p_{SI_2}) M_S - \eta_2 S\\
\dot{I_1} &=&  (r_1p_{SI_1}+r_2p_{SI_2})M_S - (\eta_2+\mu_1) I_1\\
\dot{I_2} &=&  \mu_1 I_1 - (\eta_2+\mu_2) I_2.
\end{eqnarray}
To derive the probabilities $p_{AB}$ for links from $A$ to point to $B$ we again use
\begin{equation}
\dot{p}_{AB}=\frac{\dot{M}_{AB}}{M_A}-\frac{\dot{M}_{A}}{M_A}p_{AB}
\end{equation}
with $A,B\in\{S,I_1,I_2\}$.
Analogous to \cite{ref53} and to the case of the SID model we derive that the average excess degree of a susceptible node that was reached from an infected node (either of $I_1$ or $I_2$ type) to susceptible or infected nodes are  $\delta_{SI}(I_1)=p_{SI_1}\frac{g''_S(1,t)}{g'_S(1,t)}$, $\delta_{SI}(I_2)=p_{SI_2}\frac{g''_S(1,t)}{g'_S(1,t)}$ or $\delta_{SI}(S)=(1-p_{SI_1}-p_{SI_2})\frac{g''_S(1,t)}{g'_S(1,t)}$, respectively. With this we can summarize

\begin{eqnarray}
\!\!\!\!\!\!\!\dot{M}_S&=& -(r_1p_{SI_1}+r_2p_{SI_2})S(g''_S(1,t)+g'_S(1,t))+\eta_1\bar{g}'(1,t)(N+S)-(2\eta_2+\mu_2 p_{SI_2})M_S\\
\!\!\!\!\!\!\!\dot{M}_{I_1}&=&(r_1p_{SI_1}+r_2p_{SI_2}) S(g''_S(1,t)+g'_S(1,t))+\eta_1\bar{g}'(1,t)I_1-(2\eta_2+\mu_1+\mu_2 p_{I_1I_2})M_{I_1}\\
\!\!\!\!\!\!\!\dot{M}_{I_2}&=& \mu_1 M_{I_1} +\eta_1\bar{g}'(1,t)I_2-(2\eta_2+\mu_2+\mu_2 p_{I_2I_2})M_{I_2}\\
\!\!\!\!\!\!\!&&\nonumber\\
\!\!\!\!\!\!\!\dot{M}_{SI_1}&=& -(r_1p_{SI_1}+r_2p_{SI_2})M_S (\delta_{SI}(I_1)-\delta_{SI}(S))-(r_1+\mu_1)M_{SI_1}-2\eta_2 M_{SI_1}+\eta_1\bar{g}'(1,t)I_1\\
\!\!\!\!\!\!\!&=& -(r_1p_{SI_1}+r_2p_{SI_2})(2p_{SI_1}+p_{SI_2}-1)M_S \frac{g''_S(1,t)}{g'_S(1,t)}-(r_1+\mu_1+2\eta_2)M_{SI_1}+\eta_1\bar{g}'(1,t)I_1\\
\!\!\!\!\!\!\!\dot{M}_{SI_2}&=& -(r_1p_{SI_1}+r_2p_{SI_2})M_S \delta_{SI}(I_2)-(r_2+\mu_2)M_{SI_2}+\mu_1 M_{SI_1}-2\eta_2 M_{SI_2}+\eta_1\bar{g}'(1,t)I_2\\
\!\!\!\!\!\!\!&=&-(r_1p_{SI_1}+r_2p_{SI_2})p_{SI_2}M_S \frac{g''_S(1,t)}{g'_S(1,t)}+\mu_1 M_{SI_1}-(r_2+\mu_2+2\eta_2)M_{SI_2}+\eta_1\bar{g}'(1,t)I_2\\
\!\!\!\!\!\!\!\dot{M}_{I_1I_1}&=& 2(r_1p_{SI_1}+r_2p_{SI_2})M_S \delta_{SI}(I_1)+2r_1M_{SI_1}-2\mu_1 M_{I_1I_1}-2\eta_2 M_{I_1I_1}\\
\!\!\!\!\!\!\!&=&2(r_1p_{SI_1}+r_2p_{SI_2})p_{SI_1} M_S \frac{g''_S(1,t)}{g'_S(1,t)}+2r_1M_{SI_1}-2(\mu_1+\eta_2)M_{I_1I_1}\\
\!\!\!\!\!\!\dot{M}_{I_1I_2}&=& (r_1p_{SI_1}+r_2p_{SI_2})M_S \delta_{SI}(I_2)+r_2M_{SI_2}-(\mu_1+\mu_2) M_{I_1I_2}+\mu_1 M_{I_1I_1}-2\eta_2 M_{I_1I_2}\\
\!\!\!\!\!\!\!&=& (r_1p_{SI_1}+r_2p_{SI_2})p_{SI_2} M_S \frac{g''_S(1,t)}{g'_S(1,t)}+r_2M_{SI_2}+\mu_1M_{I_1I_1} -(\mu_1+\mu_2+2\eta_2)M_{I_1I_2}\\
\!\!\!\!\!\!\dot{M}_{I_2I_2}&=&\!\! 2\mu_1 M_{I_1I_2}-2(\mu_2+\eta_2)M_{I_2I_2}
\end{eqnarray}
and write down equations for $p_{SI_1}$, $p_{SI_2}$, $p_{I_1I_1}$, $p_{I_1I_2}$ and $p_{I_2I_2}$
\begin{eqnarray}
\!\!\!\!\!\!\!\dot{p}_{SI_1}&=&(r_1p_{SI_1}+r_2p_{SI_2})(1-p_{SI_1}-p_{I_2})\frac{g''_S(1,t)}{g'_S(1,t)}+(r_1p_{SI_1}+r_2p_{SI_2})p_{SI_1}\nonumber\\
\!\!\!\!\!\!\!&&-(r_1+\mu_1-\mu_2 p_{SI_2})p_{SI_1}+\eta_1\frac{\bar{g}'(1,t)}{M_S}(I_1-(N+S)p_{SI_1})\\
\!\!\!\!\!\!\!\dot{p}_{SI_2}&=& (r_1p_{SI_1}+r_2p_{SI_2})p_{SI_2}+\mu_1p_{SI_1}-(r_2+\mu_2-\mu_2p_{SI_2})p_{SI_2}+\eta_1\frac{\bar{g}'(1,t)}{M_S}(I_2-(N+S)p_{SI_2})\\
\!\!\!\!\!\!\!\dot{p}_{I_1I_1}&=&(r_1p_{SI_1}+r_2p_{SI_2})(2p_{SI_1}-p_{I_1I_1})\frac{M_{S}}{M_{I_1}}\frac{g''_S(1,t)}{g'_S(1,t)}-(r_1p_{SI_1}+r_2p_{SI_2})\frac{M_{S}}{M_{I_1}}p_{I_1I_1}+2r_1\frac{M_{S}}{M_{I_1}}p_{SI_1}\nonumber\\
\!\!\!\!\!\!\!&&-(\mu_1-\mu_2p_{I_1I_2})p_{I_1I_1}-\eta_1\frac{\bar{g}'(1,t)}{g'_{I_1}(1,t)}p_{I_1I_1}\\
\!\!\!\!\!\!\!\dot{p}_{I_1I_2}&=&(r_1p_{SI_1}+r_2p_{SI_2})(p_{SI_2}-p_{I_1I_2})\frac{M_{S}}{M_{I_1}}\frac{g''_S(1,t)}{g'_S(1,t)}-(r_1p_{SI_1}+r_2p_{SI_2})\frac{M_{S}}{M_{I_1}}p_{I_1I_2}+r_2\frac{M_S}{M_{I_1}}p_{SI_2}\nonumber\\
\!\!\!\!\!\!\!&&+\mu_1p_{I_1I_1}-\mu_2(1-p_{I_1I_2})p_{I_1I_2}-\eta_1\frac{\bar{g}'(1,t)}{g'_{I_1}(1,t)}p_{I_1I_2}\\
\!\!\!\!\!\!\!\dot{p}_{I_2I_2}&=&\mu_1 \frac{M_{I_1}}{M_{I_2}}(2p_{I_1I_2}-p_{I_2I_2})-\mu_2(1-p_{I_2I_2})p_{I_2I_2}-\eta_1\frac{\bar{g}'(1,t)}{g'_{I_2}(1,t)}p_{I_2I_2}.
\end{eqnarray}
Again, to close the set of equations we need to derive the probability generating functions 
\begin{equation}
 \dot{g}_A(x,t) = \sum_k\left(\frac{\dot{A}_k}{A}-\frac{\dot{A}}{A}p_{Ak}\right) x^k \text{ with } A\in\{S, I_1, I_2\},
\end{equation}
to conclude with
\begin{eqnarray}
 \dot{g}_S(x,t)&=&\frac{(r_1p_{SI_1}+r_2p_{SI_2})}{S}(M_Sg_S(x,t)-xSg'_S(x,t))+\eta_1\frac{N}{S}(\bar{g}(x,t)-g_S(x,t))\nonumber\\
&&-\eta_1 (1-x)\bar{g}'(1,t)g_S(x,t)+(\eta_2+\mu_2p_{SI_2})(1-x)g'_S(x,t)\\
\dot{g}_{I_1}(x,t)&=&-\frac{(r_1p_{SI_1}+r_2p_{SI_2})}{I_1}(M_Sg_{I_1}(x,t)-xSg'_{S}(x,t))\nonumber\\
&&-\eta_1(1-x)\bar{g}'(1,t)g_{I_1}(x,t)+(\eta_2+\mu_2p_{I_1I_2})(1-x)g'_{I_1}(x,t)\\
\dot{g}_{I_2}(x,t)&=&\mu_1\frac{I_1}{I_2}(g_{I_1}(x,t)-g_{I_2}(x,t))\nonumber\\
&&-\eta_1(1-x)\bar{g}'(1,t)g_{I_2}(x,t)+(\eta_2+\mu_2p_{I_2I_2})(1-x)g'_{I_2}(x,t).
\end{eqnarray}
Note that the global quantities can be easily derived from these equations
\begin{eqnarray}
\!\!\!\!\!\!\!\!\!\dot{N}_k\!\!\!&=&\!\!\eta_1N\bar{p}_k-\eta_2 N_k- \mu_2I_{2k} +\eta_1\bar{g}'(1,t)(N_{k-1}-N_k)+\eta_2(kN_k-(k+1)N_{k+1})\nonumber\\
&&\!\!-\mu_2(p_{SI_2}(kS_k-(k+1)S_{k+1})+p_{I_1I_2}(kI_{1 k}-(k+1)I_{1 k+1})+p_{I_2I_2}(kI_{2k}-(k+1)I_{2 k+1}))\\
\!\!\!\!\!\!\!\!\!\dot{N}\!\!\!&=&\!\!(\eta_1-\eta_2) N-\mu_2 I_2\\
\!\!\!\!\!\!\!\!\!\dot{M}\!\!\!&=&\!\!2\eta_1\bar{g}'(1,t)N-2\eta_2g'(1,t)N-2\mu_2M_{I_2}\\
\!\!\!\!\!\!\!\!\!\dot{g}(x,t)\!\!\!&=&\!\!\mu_2\frac{I_2}{N}(g(x,t)-g_{I_2}(x,t))+\eta_1(\bar{g}(x,t)-g(x,t))-\eta_1(1-x)\bar{g}'(1,t)g(x,t)\nonumber\\
\!\!\!\!\!\!\!\!\!\!\!\!&&\!\!+\eta_2(1-x)g'(x,t)+\mu_2(1-x)(\frac{S}{N}p_{SI_2}g'_S(x,t)+\frac{I_1}{N}p_{I_1I_2}g'_{I_1}(x,t)+\frac{I_2}{N}p_{I_2I_2}g'_{I_2}(x,t)).
\end{eqnarray}
\subsection*{Transient contacts}

Transient contacts are considered analogous to \cite{ref54} by swapping of partners at a rate $\rho$. This affects the probabilities for links coming from $A$ nodes to point to $B$  nodes $p_{AB}$ via an additional term
\begin{equation}
\rho \left( \frac{M_B}{M}-p_{AB}\right)
\end{equation}
in the differential equations with $A,B\in\{S, I, I_1, I_2\}$.

\end{document}